\documentclass[aps,pre,twocolumn,groupedaddress,showpacs]{revtex4}
\usepackage{graphicx}

\begin{document}


\title{The shape and erosion of pebbles}


\author{D.J. Durian,$^{1,2}$ H. Bideaud,$^{2}$ P. Duringer,$^{3}$ A.
Schr\"oder,$^{2}$ C.M. Marques$^{2}$}
\affiliation{$^{1}$Department of Physics \& Astronomy, University of
Pennsylvania, Philadelphia, PA 19104-6396, USA}
\affiliation{$^{2}$LDFC-CNRS UMR 7506, 3 rue de l'Universit\'e,
67084 Strasbourg Cedex, France} \affiliation{$^{3}$CGS-CNRS UMR
7517, Institut de G\'eologie, 1 rue Blessig, 67084 Strasbourg Cedex,
France}


\date{\today}

\begin{abstract}
The shapes of flat pebbles may be characterized in terms of the
statistical distribution of curvatures measured along their
contours. We illustrate this new method for clay pebbles eroded in a
controlled laboratory apparatus, and also for naturally-occurring
rip-up clasts formed and eroded in the Mont St.-Michel bay. We find
that the curvature distribution allows finer discrimination than
traditional measures of aspect ratios. Furthermore, it connects to
the microscopic action of erosion processes that are typically
faster at protruding regions of high curvature. We discuss in detail
how the curvature may be reliable deduced from digital photographs.
\end{abstract}

\pacs{45.70.-n,83.80.Nb,91.60.-x,02.60.Jh}


\maketitle

\section{Introduction}

The roundness of pebbles on a beach has long been a source of wonder
and astonishment for scientists in many
fields~\cite{aristotle,rayleigh}.  Explanations for the pebble
shapes were born from the simple pleasure of understanding nature
but also from the hope that a pebble, or a collection of pebbles,
might carry lithographically imprinted the signature of their
erosion history. Reading that imprint would then, for instance,
reveal if a pebble was eroded on a beach, a river or a glacier, or
if it traveled a long distance down a stream.  It even perhaps would
reveal for how long the erosion forces have been at work on that
object.  Of obvious interest in Geology~\cite{boggs}, a physical
understanding of the formation of erosion shapes would also allow
for a better control of many industrial processes leading to rounded
objects such as gem stone or clay bead grinding in tumblers or fruit
and vegetable peeling in several mechanical devices. Diverse
mathematical tools have been developed for geometrical shape
analysis of crystalites, cell membranes, and other far from
equilibrium systems~\cite{chaikin,lipowsky,meakin,jamtveit};
however, these do not seem applicable to pebbles.

The evolution of a pebble shape under erosion can arguably be viewed
as a succession of elementary cuts that act at the surface of the
body to remove a given amount of material.  This converts young,
polyhedral-like shapes with a relatively small number of large sides
and sharp vertices into more mature shapes with a high number of
small sides and smooth vertices.  The size and the shapes of each of
these successive ablations, as well as the surface sites where the
cutting happens, are determined both by the conditions under which
erosion takes place and by the nature of the material being eroded.
Exposure of a young, polyhedral-like shape to the rough tumbling of
a steep stream slope will result in relatively large cuts of the
angular sections, while exposure to the gentle erosion of wind or
water is more likely to lead to small cuts almost parallel to the
existing flat sides.  Also, the same sequence of external forces
acting on two identical original shapes of different materials will
result into distinct forms due to weight, hardness or anisotropy
differences.  In spite of the diversity of factors at play in shape
modification, the complete evolution of the pebble shape is fully
determined by (i) the initial form described by some number of
faces, edges and vertices and (ii) the position, size and
orientation of the successive ablations.

Given that the erosion process evolves by a succession of localized
events on the pebble surface, it is surprising that the majority of
the precedent attempts to characterize the pebble shapes were
restricted to the determination of global quantities such as the
pebble mass or the lengths of its three main axes~\cite{boggs}.
Clearly, in order to capture both the local nature of the erosion
process and the statistical character of the successive elementary
cuts, one needs to build a new detailed description of the pebble
shapes based on quantities that are more microscopic and more
closely connected to evolution processes.  In Ref.~\cite{PebblePRL}
we proposed curvature as a key microscopic variable, since,
intuitively, protruding regions with large curvature erode faster
than flatter regions of small curvature.  We then proposed the
distribution of curvature around a flat, two-dimensional, pebble as
a new statistical tool for shape description.  And finally we
illustrated and tested these ideas by measuring and modeling the
erosion of clay pebbles in a controlled laboratory apparatus.

In this paper we elaborate on our initial Letter~\cite{PebblePRL},
and we apply our methods to naturally-occurring rip-up clasts found
in the tidal flats of the Mont St.-Michel bay.  Section~II begins
with a survey of shape quantification for two-dimensional objects,
in general, and recapitulates our new curvature-based method.
Section~III provides further details of the laboratory experiments
on clay pebbles. Section~IV presents a new field study of the Mont
St.-Michel rip-up clasts.  And finally, following the conclusion, two
methods are presented in the Appendix for reliably extracting the
local curvature from digital photographs.

\section{2d shape quantification}

The issue of rock shape is of long-standing interest in the field of
sedimentology~\cite{wentworth19,wadell32,krumbein41a,krumbein41b,
cailleux,luttig,blenk,kuenen,duringer}.  Two basic methods have
become sufficiently well established as to be discussed in
introductory textbooks~\cite{boggs}.  The simplest is a visual chart
for comparing a given rock against a standard sequence of rocks that
vary in their sphericity and angularity.  A rock has high {\it
``sphericity''} if its three dimensions are nearly equal.  And it is
{\it ``very angular''}, independent of its sphericity, if the
surface has cusps or sharp ridges; the opposite of ``very angular''
is {\it ``well rounded''}.  While useful for exposition, such verbal
distinctions are subjective and irreproducible.  The second method
is to form dimensionless shape indices based on the lengths of three
orthogonal axes.  From the ratios, and the ratios of differences, of
the long to intermediate to short axes, one can readily distinguish
rods from discs from spheres.  And a given rock may be represented
by a point on a triangular diagram according to the values of three
such indices, with rod / disc / sphere attained at the corners. This
practice is nearly half a century old~\cite{sneed}. Nevertheless,
there is still much debate about which of the infinite number of
possible shape indices are most
useful~\cite{illenberger,benn,howard,hofmann,graham}.  In any case,
such indices cannot capture fine distinctions in shape, let alone
the verbal distinctions of angular vs rounded.  Furthermore, they
provide no natural connection to the underlying physical process by
which the rock was formed.

More recent methods of shape analysis employ
Fourier~\cite{ehrlich,clark,diepenbroek,bowman}, or even
wavelet~\cite{drolon}, transforms of the contour.  This applies
naturally to flat pebbles or grains, but also to flat images of
three-dimensional objects.  The advantage of Fourier analysis over
shape indices is that, with enough terms in the series, the exact
pebble contour can be reproduced.  For simple shapes, the contour
may be described in polar coordinates by radius (eg distance from
center of mass) vs angle, $r(\theta)$, and the corresponding
transform.  However, this representation is not single-valued for
complex shapes with pits or overhangs.  Generally, the contour may
be described by Cartesian coordinates vs arclength, $\{x(s),y(s)\}$,
and the corresponding transforms.  In any case, the relative
amplitudes of different harmonics give an indication of shape in
terms of roughness at different length scales.  In a different area
of science, Fourier representations have proven especially useful
for analysis of fluctuations and instabilities of liquid interfaces,
membranes, etc.~\cite{chaikin,lipowsky,meakin}.  In practice, for
these systems, shape  fluctuations are sampled during some time
interval and then the average  Fourier amplitudes extracted by
averaging over many different realizations of the shape. Also,
because these phenomena are {\it linear}, each Fourier component
grows or shrinks at some amplitude-independent rate and the
evolution is fully determined by a dispersion relation.
Unfortunately these features do not hold for the erosion of pebbles.
Because each pebble shape only provides one configuration, average
quantities need to be built from a different prescription. Also,
there is no {\it a priori} guaranty that the variables are Gaussian
distributed, and one needs a direct space method to better assess
the importance of non-linear phenomena. Non-linearity is, we
suspect, intrinsically embedded in the  erosion mechanisms of
pebbles. If one considers for instance  a shape represented by a
single harmonic in the $r(\theta)$ representation, it is clear that
the peaks will wear more rapidly than the valleys. Therefore the
erosion rate cannot be a function of the harmonic number only; it
must either be a non-linear  function of the amplitude itself or a
function coupling many harmonics.

Our aim is to provide an alternative measure of pebble shape that is
well-defined, simple, and connects naturally to local properties
involved in the evolution process.  We restrict our attention to flat
pebbles, where an obvious shape index is the aspect ratio of long to
short axes.  Since erosion processes generally act most strongly on
the rough, pointed portions of a rock, we will focus on the local
curvature of the pebble contour.  Technically, curvature is a vector
given by ${\bf K}={\rm d}{\bf T}/{\rm d}s$, the derivative of the unit
tangent vector with respect to arclength along the
contour~\cite{weisstein}.  More intuititively, the magnitude of the
curvature is the reciprocal of the radius of a circle that mimics the
local behavior of the contour.  Here we shall adopt the sign
convention $K>0$ where the contour is convex (as at the tip of a bump)
and $K<0$ where the contour is concave (as where a chip or bite has
been removed from an otherwise round pebble).  In the Appendix, we
describe two means by which the curvature may be reliably measured at
each point along the pebble contour.  Note that the average curvature
is simply related to the perimeter of the contour:
\begin{equation}
    P=2\pi/\langle K\rangle,
\label{per}
\end{equation}
which is obviously correct when the shape is a circle.

To describe the shape of a pebble, a very natural quantity is the
distribution of curvatures, $\rho(K)$, defined such that
$\rho(K){\rm d}K$ is the probability that the curvature at some
point along the contour lies between $K$ and $K+{\rm d}K$
\cite{PebblePRL}. In order to distinguish different distributions,
as a practical matter, it is more reliable~\cite{nr} to use the
cumulative distribution of curvatures
\begin{equation}
    f(K)=\int_{0}^{K}\rho(K'){\rm d}K'
\end{equation}
Literally, $f(K)$ is the fraction of the perimeter with curvature
{\it less} than $K$.  Note that $f(K)$ increases from 0 to 1 as
$K:0\rightarrow\infty$; the minimum curvature is where $f(K)$ first
rises above 0, the maximum curvature is where $f(K)$ first reaches
1, and the median curvature is where $f(K)=1/2$.  Unlike for
$\rho(K)$, it is not necessary to bin the curvature data in order to
deduce $f(K)$.  Instead, just sort the curvature data from smallest
to largest and keep a running sum of the arclength segments,
normalized by perimeter.  Finally, so that the shapes of pebbles of
different {\it sizes} may be compared, it is useful to remove the
scale factor $\langle K\rangle$, which is related to the total
perimeter as noted above in Eq.~(\ref{per}).  Altogether, we thus
propose to quantify pebble shape by examining $f(K)$ as a function
of $K/\langle K\rangle$.

\begin{figure}
\includegraphics[width=3.00in]{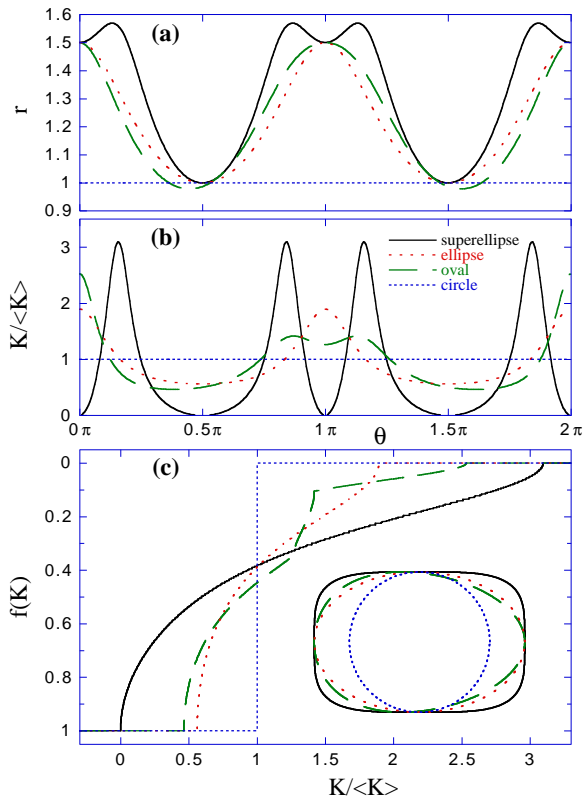}
\caption{\label{egshapes} (a) Radius, (b) normalized curvature and
(c) fraction of the perimeter $f(K)$ with curvature less than $K$,
vs $K$ divided by the average curvature $\langle K\rangle$, for a
superellipse, oval, ellipse, and circle.  The curve types match
those for the shapes as shown in the inset.  Note that, except for
the circle, all have the same aspect ratio.  The differences in
shape are reflected in differences in the forms of $f(K)$.}
\end{figure}

To help build intuition, examples of $f(K)$ are given in
Fig.~\ref{egshapes} for a few simple shapes.  The simplest of all is
a circle, where the curvature is the same at each point along the
contour: $K=\langle K\rangle = 2\pi/P = 1/R$.  Thus $f(K)=0 (1)$ for
$K< (>)\langle K\rangle$.  The curvature distribution is the
derivative of this step function, giving $\rho(K)=\delta(K-\langle
K\rangle)$ as required.  The other three shapes shown are a
superellipse, an oval, and an ellipse.  For the latter, with long
and short axes $a$ and $b$ respectively, one may compute $f(K)=
E(\sin^{-1}\lbrack 1/\epsilon (1- ( 2\sqrt{1-\epsilon^2}
E(\epsilon^2)/(\pi
K))^{2/3})^{1/2}\rbrack,\epsilon^2)/E(\epsilon^2)$ where
$\epsilon=\sqrt{1-b^{2}/a^{2}}$ is the ellipticity, $E(x)$ is the
complete elliptical integral of the first kind and $E(x,m)$ is the
incomplete elliptic integral of the second kind~\cite{weisstein}.

While noticeably different, the superellipse, ellipse, and oval in
Fig.~\ref{egshapes} all have the same aspect ratio of 1.5.  This
emphasizes how a single number is insufficient to quantify shape.  The
shape differences do, however, show up nicely in the forms of $f(K)$.
The ellipse is closest to a circle, with a distribution of curvatures
that is most narrowly distributed around the average and hence with an
$f(K)$ that is most like a step function.  The superellipse is
farthest from a cirlce, with four long nearly-flat sections and four
high-curvature corners; its curvature distribution is broadest.  The
oval is intermediate.

\section{Laboratory experiment}

The examples given in Fig.~\ref{egshapes} correspond to regular,
highly symetric shapes of two dimensional convex ``pebbles''.  In
practice, natural or artificial erosion processes lead to curvature
functions with an important statistical component.  In this section
we elaborate on the laboratory experiments of Ref.~\cite{PebblePRL},
designed to study both the statistical nature of the curvature
distribution and the influence of the original shapes of the pebbles
on the final output of a controlled erosion process.

Laboratory pebbles were formed from ``chamotte'' clay, a kind of
clay made from Kaolin and purchased from Graphigro, France.  The
water content of the purchased clay was $22$\% in a state that could
easily be kneaded.  The clay was kept tightly packed before use in
order to avoid water evaporation.  Clay pebbles were produced using
aluminium molds made in our laboratory. They consist of a polygonal
well of $0.5$ cm depth.  Once the mould was filled with clay, it was
left at rest for 24 hours, so that $98.5$\% of the water was removed
by evaporation.  All the experimental results presented here concern
one day old pebbles.  We noticed that pebbles older than 2 days were
too fragile for our experimental configuration. The dimensions of
the various pebbles studied here are as follows: $4$ squares of side
$5$~cm, $5$ rectangles of sides $4\times 6$~cm$^{2}$, $5$ regular
pentagons of side $4.25$~cm, one triangle with sides $7$~cm,$7.5$~cm
and $9$~cm one irregular polygon with $7$ sides, one lozenge of
acute angles $45^\circ$ and side $5$~cm and one circle of diameter
$7$~cm.

\begin{figure}
\includegraphics[width=3.00in]{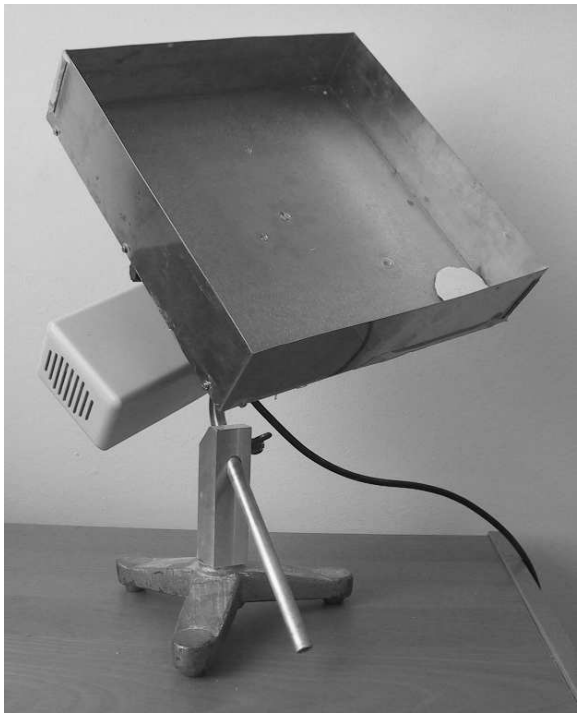}
\caption{
The wearing apparatus used for the laboratory
experiments.  The rotating metal tray is $30\times 30\times 7$~cm$^{3}$.}
\label{plateau}
\end{figure}

The wearing method that we chose relies on placing a pebble in the
rotating apparatus sketched in Fig.~\ref{plateau}.  The apparatus is
a square basin, of dimensions $30\times 30\times 7$~cm$^3$.  The
basin bottom is a $1$~cm thick aluminium plate and the walls are
made of $0.04$~cm thick aluminium sheets.  This rotating plate is
fixed to a rod held by the jaws of a laboratory mixer, Heidolph
RZR1.  The mixer itself is fixed to a tripod, so that its
inclination angle can be varied.

A typical trajectory of the pebble during the continuous rotation of
the plate can be described as follows.  First the pebble rotates
with the basin until it reaches a high position.  After the plate
has rotated an angle between $\pi/2$ and $\pi$, the pebble begins to
slide due to gravity, until it hits one of the walls in the bottom
part of the container, and then rolls down along that wall as the
basin keeps its rotation.  After a short stop at a container corner
the pebble starts a new cycle again.  We performed preliminary tests
in order to determine both ideal basin orientation and ideal
rotation frequency for our experiments.  As expected, above some
maximum rotation frequency the pebble becomes immobilised in the
basin: centrifugal forces maintain the pebble on a given position
against the wall. Also, under some minimum inclination angle, no
fall of the pebble is observed, while a high inclination doesn't
allow the pebble to reach it's maximum altitude.  Altogether, we
found it suitable to operate at a basin angle of $45^\circ$ and a
rotation frequency of one cycle per second.  Using the latter
experimental conditions, we observed that the in-plane dimension of
a pebble decreased by around a factor 2 after 30 minutes.  Thus, a
significant wearing of a pebble could be observed after a few
minutes rotation.  In practice, each pebble was eroded under the
described conditions during 30 minutes, while a picture of the
pebble was taken after each 5 minutes wearing.  Hence, for each of
the pebbles studied, we obtained about 7 pictures, corresponding to
the initial pebble and to six following states of the wearing
pebble. For some of the pebbles 8 or 9 pictures at 5 minutes
interval were taken.  The images were then analyzed following the
method described in the appendix.

\begin{figure}
\includegraphics[width=3.00in]{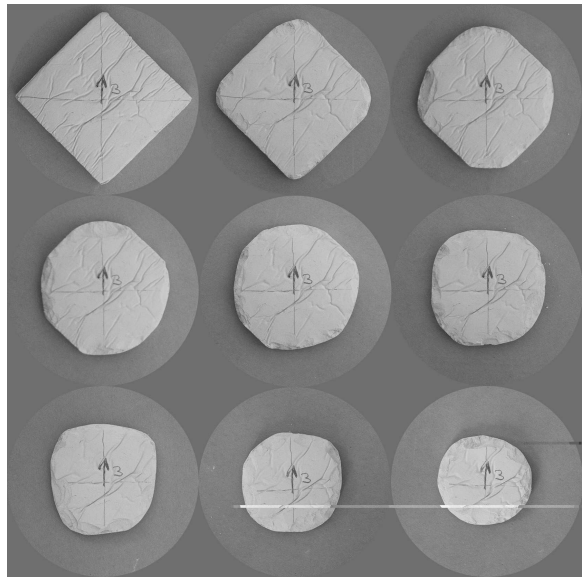}
\caption{\label{shapevol} Shape evolution of a $5\times 5$~cm square
pebble eroded in the laboratory, by the method explained in the text.}
\end{figure}

\begin{figure}
\includegraphics[width=3.00in]{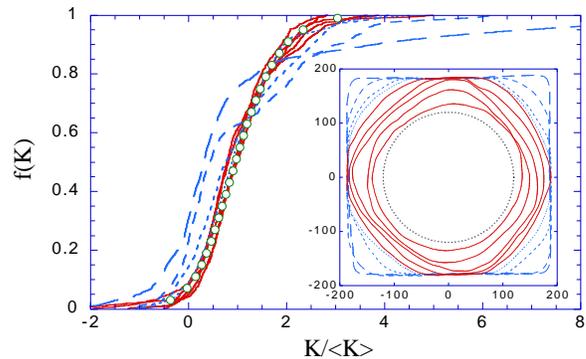}
\caption{\label{c3icds} Cumulative curvature distribution, $f(K)$, for
the evolving pebble depicted in the inset (and pictured in
Fig.~3).  As the pebble becomes
progressively rounder, the curvature distribution narrows and
approaches a final shape.  The time interval between successive
contours is 5 minutes.  For the inset, the axes are given in pixel
units, equal to 0.132 mm.  For contrast, a circle is shown by points.}
\end{figure}

An example for the shape evolution produced by this method is given in
Fig.~\ref{shapevol}, with photographs shown every five minutes.  The
corresponding cumulative curvature distributions $f(K)$ are given in
Fig.~\ref{c3icds}, where the inset shows the extracted contours.  Here
the initial shape is square, with four long nearly-flat regions and
four short high-curvature regions.  Thus the initial $f(K)$ rises
steeply around $K=0$ and extends with relatively little weight out to
$K\gg \langle K\rangle$.  At first, the action of erosion is most
rapid at the high-curvature corners, with the flat regions in between
relatively unaffected.  Thus the high-$K$ tail of $f(K)$ at first is
suppressed, and weight builds up across $(0.5-2)\langle K\rangle$.
Next the rounded corners erode further and gradually extend across the
flat sections.  Thus weight in $f(K)$ is gradually concentrated more
and more toward $\langle K\rangle$.  After about 15-20 minutes, when
the flat sections are nearly gone, the form of $f(K)$ fluctuates
slightly but ceases to change in any systematic manner.  In
otherwords, the shape of the pebble has reached a final limiting form.
{\it Further erosion will affect pebble size, but not pebble shape!}

\begin{figure}
\includegraphics[width=3.00in]{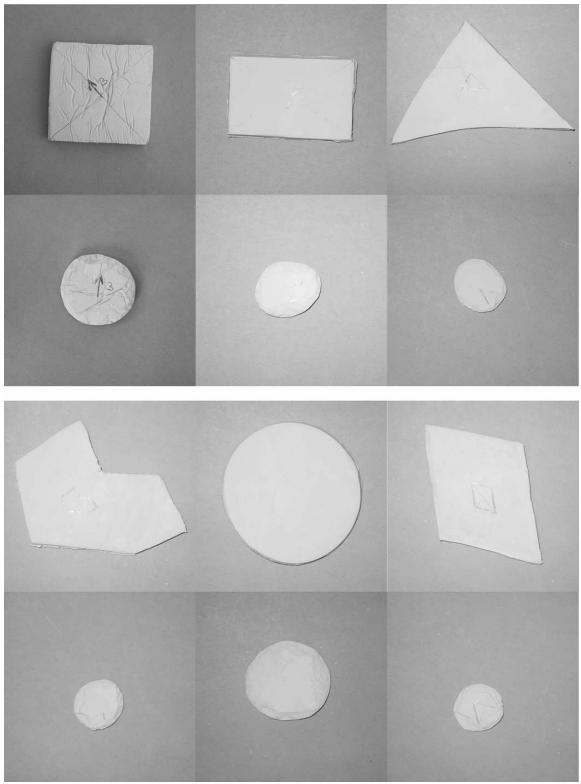}
\caption{\label{allshapes} The initial and final forms of different
shapes eroded in our experiment.  Final shapes for all cases are not
only similar in shape and in size but they have also the same
curvature distribution.}
\end{figure}

\begin{figure}
\includegraphics[width=3.00in]{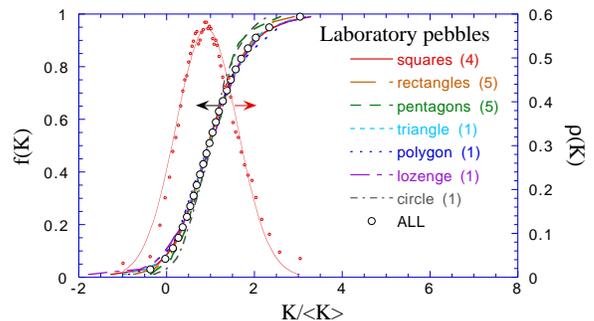}
\caption{\label{labave} Integrated curvature distribution, $f(K)$, for
the final shapes of laboratory pebbles with various initial shapes, as
labeled.  These are indistinguishable to within measurement accuracy;
their average is shown by the open circles.  The corresponding average
curvature distribution, $\rho(K)$ is obtained by differentiation and
is shown on the right axis along with a fit to a Gaussian shape.}
\end{figure}

To test the universality of the final shape, we repeat the same
experiment both for other squares as well as for a variety of other
initial shapes such as rectangles, triangles, and circles.  A number
of different examples showing both the initial and final shapes are
shown in Fig.~\ref{allshapes}.  In all cases, the cumulative curvature
distribution $f(K)$ shows a systematic evolution at short times and
slight fluctuations about some average shape at later times, just as
in Fig.~\ref{c3icds}.  The more angular or oblong the initial shape,
the more erosion is needed to reach a stationary final shape.  The
average final $f(K)$ is shown for the various initial shapes in
Fig.~\ref{labave}.  Evidently, these all display the same quantitative
form independent of the initial shape.  Even $f(K)$ for an
initially-circular pebble broadens from a step function to the same
form as all the others.

The final $f(K)$ for all initial shapes can thus be averaged
together for a more accurate description of the stationary shape
produced by the laboratory erosion machine.  The result is shown by
the open circles in the same plot, Fig.~\ref{labave}.
Differentiating, we obtain the actual curvature distribution,
$\rho(K)$, shown on the right axis.  It is fairly broad, with a
full-width at half-maximum equal to about $1.6\langle K\rangle$. The
actual shape is not quite symmetrical, skewed toward higher
curvatures.  The closest simple analytic form would be a Gaussian,
$\exp[-(K-\langle K\rangle)^{2}/(2\sigma^{2})]$.  The actual
distribution is slightly skewed toward higher curvatures, but the
best fit gives a standard deviation of $\sigma=0.70\langle
K\rangle$, as shown in Fig.~\ref{labave}.  It is easy to imagine
that the width of this distribution could be set by the strength of
the erosion process.  For example, if the angle of the rotating pan
were lowered, then the erosion would be more gradual and more like
polishing; in which case a rounder stationary shape may be attained
with a narrower distribution of curvatures.  The form of $f(K)$, as
well as its width, could also be affected.  These types of questions
can be addressed, both in laboratory and field studies, now that we
have an incisive tool like $f(K)$ for quantifying shape.

\begin{figure}
\includegraphics[width=3.00in]{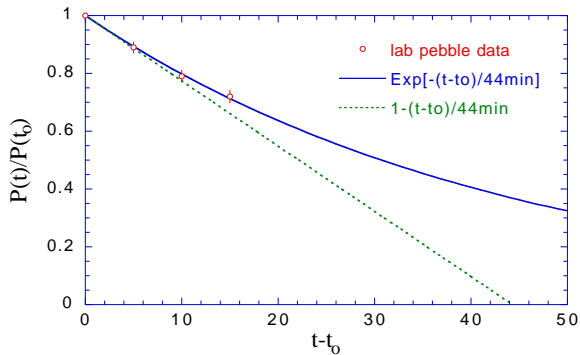}
\caption{\label{perim} Perimeter vs time, where $t_{0}$ is when the
stationary shape has been reached.}
\end{figure}

To further study the erosion produced by our laboratory apparatus,
we now consider how the perimeter of the pebble decreases with time,
$P(t)$.  Since the initial behavior depends on the specific initial
shape, we focus on subsequent erosion once the (universal)
stationary shape is achieved.  If the final stationary shape of the
curvature distribution is reached at time $t_{0}$, then the quantity
of interest is really $P(t)/P(t_{0})$ vs $t-t_{0}$.  The results,
averaged over all laboratory pebbles, are shown in Fig.~\ref{perim}.
Though the dynamic range is not great, the data are consistent with
an exponential decrease, $P(t)=P(t_{0})\exp[-(t-t_{0})/\tau]$.  The
best fit to this form is shown by a solid curve; it gives a decay
constant of $\tau=44$~min.  Exponential erosion is, in fact,
observed in field and laboratory studies~\cite{krumbein41b}.  It is
to be expected whenever the strength of the erosion is proportional
to the pebble size, as in our lab experiments where the impulse upon
collision is proportional to the pebble's weight.

\section{Field study}

As a first field test of our method of analysis, we collected mud
pebbles in the Mont St.-Michel bay, France.  The littoral
environment located at the inner part of the Norman-Breton Gulf is
characterized by a macro-tidal dynamics.  This location exhibits the
second largest tide in the world after the Bay of Fundy, Canada.
During the spring tide periods the upper part of the tidal flats
collects a muddy sediment.  This mud dries up during the following
neap tide period where the sediments are exposed to the air. In
certain areas, between the large equinoxial neap tides, the exposure
of mud sediments to air may last for several months. During this
period, this mud layer will develop a vast network of desiccation
cracks. This network then leads to fragmented plates of a polygonal
shape with 20 to 40~cm size.  During the next spring tide period,
the plates in the erosional area can be eroded by tide currents,
thus re-incorporated into the sedimentary cycle.  During the tide
process these clasts are progressively eroded over many months. Thus
the mud cohesion allows enough observation time for the life of a
clast to be observed within a distance of order of one hectometer,
from the original erosional area down to the latest stages of
abrasion.

\begin{figure}
\includegraphics[width=3.00in]{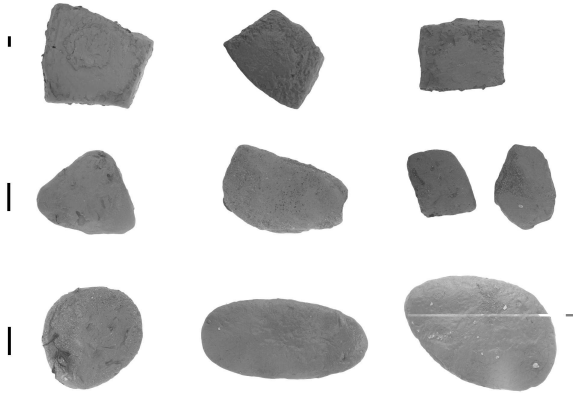}
\caption{\label{figterrain} Typical shapes of Mont St.-Michel rip-up
clasts.  The immature pebbles in the top row were collected close to
their origin; the sub-mature pebbles in the middle row were collected
further downstream; the smooth, mature pebbles in the bottom row were
collected on a nearby sandbar.  As these pebbles eroded, their shapes
became rounder, an effect quantified in the next figure.  The bars
indicate, in each row, a length of 2 cm.}
\end{figure}

We have analyzed the shapes of three classes of rip-up clast
photographed at three distinct locations on the tidal flats near Mont
St.-Michel.  The first is large sub-angular cobble found near the site of
formation.  Ten samples were examined; for these immature pebbles, the
average perimeter is 550 mm.  The second class is medium sub-mature
pebbles found further ``downstream''.  As a result of erosion, these
pebbles are smaller and smoother than the cobble.  Thirty five samples
were examined; for these, the average perimeter is 180 mm.  The third
class is rounded, mature pebbles found on a nearby sand bar.  The relation
of these pebbles to the other two classes is not clear.  Seventeen samples
were examined; for these, the average perimeter is 220 mm.

\begin{figure}
\includegraphics[width=3.00in]{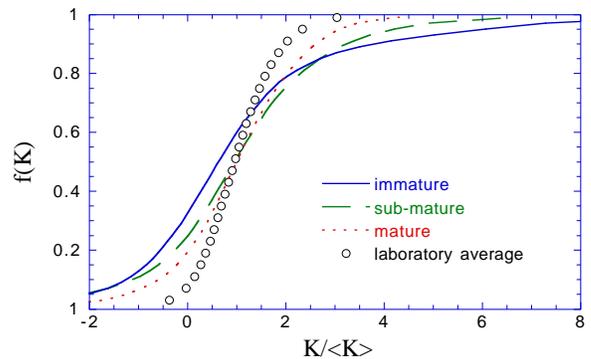}
\caption{\label{fieldave} Cumulative curvature distribution, $f(K)$,
for the average shapes of Mont St.-Michel rip-up clasts.  Even the
roundest shapes remained less circular than the final shape in the
laboratory study.}
\end{figure}

Typical photographs for each of these classes are shown in
Fig.~\ref{figterrain}.  The average of the cumulative curvature
distribution for all samples in each class is shown in
Fig.~\ref{fieldave}.  The angularity of the large cobble is reflected
in the breadth of the curvature distribution.  Roughly a quarter of
the perimeter has negative curvature, and roughly a tenth has
curvature five times greater than the average.  For the other two
classes, the curvature distribution is progressively more narrow.  The
relative steepness of the $f(K)$'s shows that all of these shapes are
less round than the final pebbles produced by the laboratory erosion
machine.

The width of the curvature distribution can be specified
quantitatively by the standard deviation, $\sigma$.  Results are
normalized by the average curvature, and are shown for the field and
laboratory pebbles in Table~\ref{curvtable}.  The pebbles with steeper
$f(K)$ indeed have smaller widths.  For example, the width for the
immature field cobble is about 3-4 times that of the average laboratory
pebble.  While the dimensionless width of the distribution,
$\sigma/\langle K\rangle$, is a useful number for comparisons, it does
not distinguish between curvature distributions of different shape.
The actual functional form of the curvature distribution can be
specified to some extent by comparing its moments with that of a
Gaussian.  In particular, the ``skewness'' and ``kurtosis'' are
dimensionless numbers defined by the third and fourth moments,
respectively, in such a way as to vanish for a perfect Gaussian.  The
results in Table~\ref{curvtable} show that the four classes of pebbles
have curvature distributions of four distinct forms.  Of these, the
laboratory pebbles are closest to a Gaussian.

\begin{table}[ht]
\begin{ruledtabular}
\begin{tabular}{lcccc}
Class & Perimeter (mm) & $\sigma/\langle K\rangle$ & Skewness & Kurtosis \\
\colrule
immature & 550$\pm100$ & 2.4$\pm0.3$ & 1.4$\pm0.3$ & 3.8$\pm1.4$ \\
sub-mature & 180$\pm60$ & 2.0$\pm0.6$ & 0.0$\pm0.9$ & 3.3$\pm2.9$ \\
mature & 220$\pm70$ & 1.4$\pm0.3$ & -0.2$\pm0.5$ & 1.3$\pm1.0$ \\
lab-final & 122$\pm25$ & 0.8$\pm0.1$ & 0.1$\pm0.5$ & 1.0$\pm1.5$ \\
\end{tabular}
\end{ruledtabular}
\caption{\label{curvtable} Characteristics of curvature distribution for
the three classes of field pebbles.  Final laboratory pebble shape is
added for comparison.}
\end{table}

\section{Conclusion}

We have studied the formation of two-dimensional pebble shapes.  As
in Ref.~\cite{PebblePRL} we introduced a local description of the
erosion process, based on the distribution function of the
curvature, measured along the pebble contours.  This description
captures both the local character of the erosion events, and the
statistical nature of the erosion process.

For pebbles generated in the laboratory, we have shown that the
curvature distribution has two important properties.  First, the
erosion drives the distribution towards a stationary form.  When this
stationary state is reached, the pebble contour still changes but,
within small fluctuations, its curvature distribution remains the
same, provided that the curvature is normalized by its average value.
Secondly, we have found that the final stationary form of the
distribution is independent of the original pebble shapes.  This not
only shows that the curvature distribution is a property of the
erosion process itself, but it also opens the interesting possibility
of establishing a classification of different erosion processes
according to the type of curvature distribution they generate.

For pebbles collected in the field, we have made a first attempt to
study a special class of rip-up clasts from the St.-Michel bay.  These
mud pebbles can be collected at very different erosion stages within a
relatively small area of the tidal flats.  We showed that the
curvature distribution sharpens with the wearing degree, without
getting however as sharp as the distribution obtained in the
laboratory experiments.

The results presented in this experimental paper suggest a number of
directions for modeling the formation of flat pebbles. Of central
importance is the intrinsic statistical nature of the erosion
process itself. As first hinted in Ref.~\cite{PebblePRL}, a sequence
of cuts of a noiseless, deterministic nature typically leads to a
trivial curvature distribution like that of a circle. We also
demonstrated in Ref.~\cite{PebblePRL} that a ``cutting'' simulation,
with an appropriate distribution of cutting lengths, acting most
strongly on regions of high curvature in accord with Aristotle's
intuition~\cite{aristotle}, can reproduce the curvature distribution
from the laboratory experiments. We will address these and other
questions relevant for the theoretical modeling of pebble formation
in a forthcoming paper.

\begin{acknowledgments}
We acknowledge insightful discussions with F. Thalmann and
experimental insight by P. Boltenhagen.  This work was supported by
the Chemistry Department of the CNRS, under AIP ``Soutien aux Jeunes
Equipes'' (CM).  It was also supported by the National Science
Foundation under Grant DMR-0514705 (DJD).
\end{acknowledgments}

\appendix*
\section{Curvature analysis}

\begin{figure}
\includegraphics[width=3.00in]{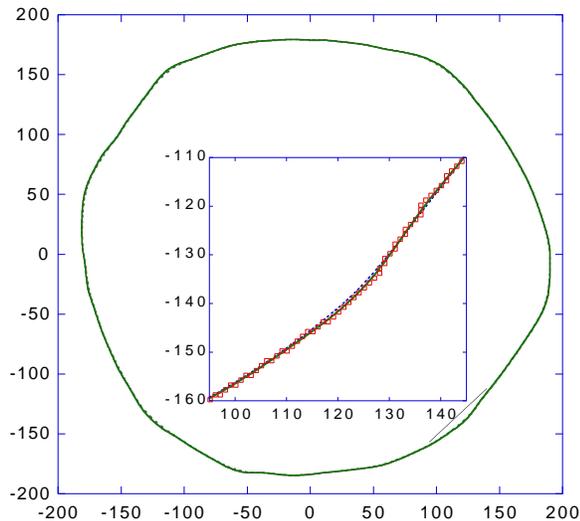}
\caption{\label{reconst} Reconstruction of a smooth pebble contour
from the pixelized digital representation.  The solid curve is based
on fitting to cubic polynomials at each point.  The dashed curve is
based on the iteration scheme of Fig.~\protect\ref{iter}.}
\end{figure}

The goal of this section is to provide a detailed, practical
description of two means to measure the local curvature at each
point along the pebble contour.  In both cases the starting point is
an image of the pebble.  For our work we use a digital camera Canon
Power Shot G1 with a resolution of $1024\times768$ pixels.  It
should be just as effective to scan conventional photographs, or
even to scan pebbles themselves.  To determine the \{x,y\}
coordinates of the contour, we import the images into NIH
Image~\cite{nih}, which includes routines for finding edges and for
skeletonizing the result.  An example of the digitized pebble
contour, and the smooth reconstructions to be discussed below, is
given in Fig.~\ref{reconst}.  We show the contour in pixel units,
since the actual calibration is needed only to determine size, not
{\it shape}.  Note from the inset that the contour points are indeed
pixelized and skeletonized, with each point having only two
neighbors located at either $\pm1$ or 0 units away in the x- and
y-directions.  If the digitization process is faithful, then the
uncertainty in each pixel is about $\pm0.5$ units in each direction.
This is not small compared to the distance between neighboring
pixels, so a smoothing or fitting routine is necessary to
reconstruct the actual contour and thereby extract the curvature
distribution.

To illustrate the difficulties of extracting curvature, let us begin
with two methods that are, in fact, unsatisfactory.  It is perhaps
tempting to simply smooth the data, replacing each point with a
weighted average of neighbors lying within some window.  Weights
could be cleverly chosen to de-emphasize points at the edge of the
window, for example.  This fails, however, since it's far from
obvious how to choose a suitable window size.  For instance, the
pixelized representation of the straight section given by $y=0.1x$
for $0<x<10$ is a step function $y=0(1)$ for $x<(>)5$.  A large
window would be needed to even approximately reconstruct the
original line.  However, such a sufficiently large smoothing window
would erase fine features if applied elsewhere along the contour.
Since smoothing filters provide no feedback on quality, visual
inspection of the result would be necessary to choose an optimal
window size at each point along the contour.  This is not only
subjective, but rather impractical.  As an alternative, it is
perhaps tempting to implement an automated version of Wentworth's
curvature gauge~\cite{wentworth19}.  This is a device with circular
notches of various diameters into which portions of a pebble may be
pressed.  The computational analogue would be to find the best
nonlinear-least-squares fit to a circle at each point along the
pebble contour.  As with smoothing, one difficulty is to find the
optimal window over which to do the fitting.  A compounding
difficulty is that essentially nowhere is the pebble exactly
circular, so even with `the' optimal window there is substantial
disagreement between data and fit.  A spectacular example of this
problem is at an inflection point, where the curvature changes from
positive to negative.

To overcome such difficulties, we propose to fit digitized contour
data to a cubic polynomial at each point along the contour.  A cubic
is the lowest order polynomial needed in order to avoid systematic
error when the curvature varies gradually across the fitting window,
which is the usual case.  In order to avoid having to rotate the
coordinate system to ensure that the contour $y(x)$ is a single-valued
function, we instead convert to polar coordinates.  Thus we define the
origin by the center-of-mass of the contour and perform fits to
$r(\theta)$ where $r=\sqrt{x^{2}+y^{2}}$ and $\theta=\tan^{-1}(y/x)$.
Once a satisfactory fit is achieved, the curvature may be deduced from
the value and derivatives of the cubic polynomial by
\begin{equation}
    K = {r^{2} + 2r_{\theta}^{2}-rr_{\theta\theta} \over
    (r^{2}+r_{\theta}^{2})^{3/2} },
\label{kvsr}
\end{equation}
where $r_{\theta}={\rm d}r/{\rm d}\theta$ and $r_{\theta\theta}={\rm
d}^{2}r/{\rm d}\theta^{2}$~\cite{weisstein}.

Two tricks seem necessary to achieve satisfactory results.  The
first is to weight the data most heavily near the center of the
window.  We use a Gaussian weighting function with a standard
deviation equal to $1/4$ of the width of the window.  This ensures
that points at the edges have essentially no influence.  Therefore,
the fitting results do not vary rapidly as the window is slid along
the contour.  This guarantees that the reconstructed curve and its
first two derivatives are continuous, which is a crucial requirement
for measuring the curvature.

The second trick is to choose the window size appropriately.  This
is actually the most difficult and subtle aspect of the whole
problem. If the window is too small, then the fit will reproduce the
bumps and wiggles of the pixelization process; usually the curvature
will be overestimated.  If the window is too large, then the fit
will significantly deviate from the data; usually the curvature will
be underestimated.  And while the curvature tends to decrease
systematically with window size, there is in general, unfortunately,
no plateau between these two extremes where the curvature is
relatively independent of window size and hence clearly represents
the true value.  To pick the window appropriately requires careful
understanding of the numerical fitting procedure and the feedback it
provides.  Since the fitting function is a polynomial, the
minimization of the $\chi^{2}$ total square deviation from the data
reduces to solving a set of linear equations.  This in turn reduces
to inverting a matrix.  If the window is too small then the fit will
be `ambiguous' in the sense that $\chi^{2}$ is small but the error
in the fitting parameters is large.  Mathematically, the matrix to
be inverted is essentially singular.  A good strategy is therefore
to start with a small window and increase its size until the matrix
is no longer singular.  This can be accomplished using a linear
least-squares fitting routine based on singular-value
decomposition~\cite{nr}. However, the uncertainty in fitting
parameters for the first suitable window is generally too large
(nearly $100\%$).  So we increase the window size two pixels at a
time until the error in curvature has been reduced by a factor of
ten.  This defines the largest suitable window, beyond which
systematic errors due to incorrect functional form begin to appear.
We have not been able to define the largest suitable window based on
the value of $\chi^{2}$.  For the final result, we take a weighted
average of the cubic fit parameters over all suitable windows, where
the weights are set by the uncertainties in fitting parameters as
returned by the fitting routine.  An example of a reconstructed
contour from this procedure is shown by the solid curve in
Fig.~\ref{reconst}.  The recontruction is satisfactorily smooth;
also, it clearly avoids the pixel noise without smoothing over
significant small-scale features in the contour.

Since the cubic fitting method is rather involved, and since the
choice of window sizes is still slightly subjective, we have
developed an alternative method.  The starting point is the fact
that the actual digital representation of the contour depends on the
location and orientation of the pebble with respect to the grid of
pixels.  If the pebble were shifted or rotated, then the pixelized
representation would be slightly different.  For example, imagine
the pixelization of a line making various angles with the grid.
Perhaps the ideal experimental measurement procedure would be to
systematically reposition the pebble, pixelize, then compute the
average of all such representations.  However this procedure does
not lend itself to automatization, and would be impractically
time-consuming. Instead, we propose to do more or less the same
thing numerically. The idea is to take the current best guess for
the contour, pixelize it with respect to a random grid position and
orientation, then use the new representation to update the best
guess.  When iterated, this procedure converges to a satisfactory
reconstruction of the actual contour with two provisos.  First, at
each step, we locally smooth the trial pixelization by replacing
each point by its average with its two immediate neighbors.  Second,
we keep only every fourth or fifth point in the original pixelized
data and perform all operations on this subset.  When done, we
compute the curvature literally by the change of slope with respect
to arclength using the straight segments between adjacent points.

\begin{figure}
\includegraphics[width=3.00in]{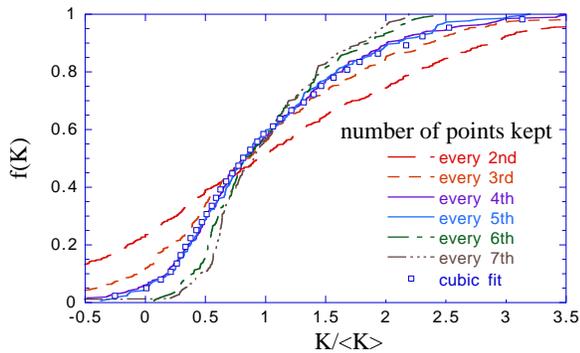}
\caption{\label{iter} Trial integrated curvature distributions from
the iterative smoothing scheme, vs the number of points kept.  Keeping
either 4 or 5 points seems optimal: the resulting distributions are
identical and they agree with that based on cubic fits.}
\end{figure}

The cumulative curvature distribution given by this iteration scheme
is shown in Fig.~\ref{iter} as a function of the number of points
kept.  When too many are kept, the reconstructed curve follows the
bumps and wiggles of the original pixelization too closely; the
curvature distribution is too broad.  When too few are kept, the
reconstructed curve incorrectly smooths over small-scale features;
the curvature distribution is too narrow.  When only every fourth or
fifth point are kept, the distributions are nearly equal;
furthermore, they are indistinguishable from that given by the cubic
polynomial fitting. The actual reconstructed contour is also shown
in Fig.~\ref{reconst}. The plateau in the curvature distribution vs
number of points kept, and the good agreement with the other method,
both give confidence in this new iterative reconstruction scheme.

\bibliography{PebRefs}

\begin{thebibliography}{31}
\expandafter\ifx\csname natexlab\endcsname\relax\def\natexlab#1{#1}\fi
\expandafter\ifx\csname bibnamefont\endcsname\relax
  \def\bibnamefont#1{#1}\fi
\expandafter\ifx\csname bibfnamefont\endcsname\relax
  \def\bibfnamefont#1{#1}\fi
\expandafter\ifx\csname citenamefont\endcsname\relax
  \def\citenamefont#1{#1}\fi
\expandafter\ifx\csname url\endcsname\relax
  \def\url#1{\texttt{#1}}\fi
\expandafter\ifx\csname urlprefix\endcsname\relax\def\urlprefix{URL }\fi
\providecommand{\bibinfo}[2]{#2}
\providecommand{\eprint}[2][]{\url{#2}}

\bibitem[{\citenamefont{Aristotle}(2000)}]{aristotle}
\bibinfo{author}{\bibnamefont{Aristotle}}, \emph{\bibinfo{title}{Minor Works,
  Mechanical Problems}} (\bibinfo{publisher}{Harvard University Press},
  \bibinfo{address}{Cambridge}, \bibinfo{year}{2000}),
  \bibinfo{note}{translation by W.S. Hett. Question 15 asks: ``Why are the
  stones on the seashore which are called pebbles round, when they are
  originally made from long stones and shells? Surely it is because in movement
  what is further from the middle moves more rapidly. For the middle is the
  center, and the distance from this is the radius. And from an equal movement
  the greater radius describes a greater circle. But that which travels a
  greater distance in an equal time describes a greater circle. Things
  travelling with a greater velocity over a greater distance strike harder, and
  things which strike harder are themselves struck harder. So that the parts
  further from the middle must always get worn down. As this happens to them
  they become round. In the case of pebbles, owing to the movement of the sea
  and the fact that they are moving with the sea, they are perpetually in
  motion and are liable to friction as they roll. But this must occur most of
  all at their extremities.''}.

\bibitem[{\citenamefont{{Lord Rayleigh, F.R.S.}}(1944)}]{rayleigh}
\bibinfo{author}{\bibnamefont{{Lord Rayleigh, F.R.S.}}},
  \bibinfo{journal}{Nature} \textbf{\bibinfo{volume}{3901}},
  \bibinfo{pages}{169} (\bibinfo{year}{1944}).

\bibitem[{\citenamefont{{S. Boggs, Jr.}}(2001)}]{boggs}
\bibinfo{author}{\bibnamefont{{S. Boggs, Jr.}}},
  \emph{\bibinfo{title}{Principles of sedimentology and stratigraphy}}
  (\bibinfo{publisher}{Prentice Hall}, \bibinfo{address}{N.J.},
  \bibinfo{year}{2001}), \bibinfo{edition}{3rd} ed.

\bibitem[{\citenamefont{Chaikin and Lubensky}(1995)}]{chaikin}
\bibinfo{author}{\bibfnamefont{P.~M.} \bibnamefont{Chaikin}} \bibnamefont{and}
  \bibinfo{author}{\bibfnamefont{T.~C.} \bibnamefont{Lubensky}},
  \emph{\bibinfo{title}{Principles of Condensed Matter Physics}}
  (\bibinfo{publisher}{Cambridge University Press}, \bibinfo{address}{New
  York}, \bibinfo{year}{1995}).

\bibitem[{\citenamefont{Lipowsky and Sackmann}(1995)}]{lipowsky}
\bibinfo{author}{\bibfnamefont{R.}~\bibnamefont{Lipowsky}} \bibnamefont{and}
  \bibinfo{author}{\bibfnamefont{E.}~\bibnamefont{Sackmann}},
  \emph{\bibinfo{title}{Structure and dynamics of membranes}}, Handbook of
  biological physics (\bibinfo{publisher}{Elsevier},
  \bibinfo{address}{Amsterdam}, \bibinfo{year}{1995}).

\bibitem[{\citenamefont{Meakin}(1998)}]{meakin}
\bibinfo{author}{\bibfnamefont{P.}~\bibnamefont{Meakin}},
  \emph{\bibinfo{title}{Fractals, scaling and growth far from equilibrium}}
  (\bibinfo{publisher}{Cambridge University Press},
  \bibinfo{address}{Cambridge}, \bibinfo{year}{1998}).

\bibitem[{\citenamefont{Jamtveit and Meakin}(1999)}]{jamtveit}
\bibinfo{editor}{\bibfnamefont{B.}~\bibnamefont{Jamtveit}} \bibnamefont{and}
  \bibinfo{editor}{\bibfnamefont{P.}~\bibnamefont{Meakin}}, eds.,
  \emph{\bibinfo{title}{Growth, dissolution and pattern formation in
  geosystems}} (\bibinfo{publisher}{Kluwer Academic Publishers},
  \bibinfo{address}{Dordrecht}, \bibinfo{year}{1999}).

\bibitem[{\citenamefont{Durian et~al.}(2006)\citenamefont{Durian, Bideaud,
  Duringer, Schroder, Thalmann, and Marques}}]{PebblePRL}
\bibinfo{author}{\bibfnamefont{D.}~\bibnamefont{Durian}},
  \bibinfo{author}{\bibfnamefont{H.}~\bibnamefont{Bideaud}},
  \bibinfo{author}{\bibfnamefont{P.}~\bibnamefont{Duringer}},
  \bibinfo{author}{\bibfnamefont{A.}~\bibnamefont{Schroder}},
  \bibinfo{author}{\bibfnamefont{F.}~\bibnamefont{Thalmann}}, \bibnamefont{and}
  \bibinfo{author}{\bibfnamefont{C.}~\bibnamefont{Marques}},
  \bibinfo{journal}{Phys. Rev. Lett} p. \bibinfo{pages}{to appear}
  (\bibinfo{year}{2006}).

\bibitem[{\citenamefont{Wentworth}(1919)}]{wentworth19}
\bibinfo{author}{\bibfnamefont{C.~K.} \bibnamefont{Wentworth}},
  \bibinfo{journal}{Journal of Geology} \textbf{\bibinfo{volume}{27}},
  \bibinfo{pages}{507} (\bibinfo{year}{1919}).

\bibitem[{\citenamefont{Wadell}(1932)}]{wadell32}
\bibinfo{author}{\bibfnamefont{H.}~\bibnamefont{Wadell}},
  \bibinfo{journal}{Journal of Geology} \textbf{\bibinfo{volume}{40}},
  \bibinfo{pages}{443} (\bibinfo{year}{1932}).

\bibitem[{\citenamefont{Krumbein}(1941{\natexlab{a}})}]{krumbein41a}
\bibinfo{author}{\bibfnamefont{W.~C.} \bibnamefont{Krumbein}},
  \bibinfo{journal}{Journal of Sedimentary Petrology}
  \textbf{\bibinfo{volume}{11}}, \bibinfo{pages}{64}
  (\bibinfo{year}{1941}{\natexlab{a}}).

\bibitem[{\citenamefont{Krumbein}(1941{\natexlab{b}})}]{krumbein41b}
\bibinfo{author}{\bibfnamefont{W.~C.} \bibnamefont{Krumbein}},
  \bibinfo{journal}{Journal of Geology} \textbf{\bibinfo{volume}{49}},
  \bibinfo{pages}{482} (\bibinfo{year}{1941}{\natexlab{b}}).

\bibitem[{\citenamefont{Cailleux}(1947)}]{cailleux}
\bibinfo{author}{\bibfnamefont{A.}~\bibnamefont{Cailleux}},
  \bibinfo{journal}{C. R. Soc. Geol. Fr. Paris} \textbf{\bibinfo{volume}{13}},
  \bibinfo{pages}{250} (\bibinfo{year}{1947}).

\bibitem[{\citenamefont{Luttig}(1956)}]{luttig}
\bibinfo{author}{\bibfnamefont{G.~G.} \bibnamefont{Luttig}},
  \bibinfo{journal}{Eizeitalter u. Gegenwart (Ohningen)}
  \textbf{\bibinfo{volume}{7}}, \bibinfo{pages}{13} (\bibinfo{year}{1956}).

\bibitem[{\citenamefont{Blenk}(1960)}]{blenk}
\bibinfo{author}{\bibfnamefont{M.}~\bibnamefont{Blenk}}, \bibinfo{journal}{Z.
  Geomorph. Berlin NF 4} \textbf{\bibinfo{volume}{3/4}}, \bibinfo{pages}{202}
  (\bibinfo{year}{1960}).

\bibitem[{\citenamefont{Kuenen}(1965)}]{kuenen}
\bibinfo{author}{\bibfnamefont{P.~H.} \bibnamefont{Kuenen}},
  \bibinfo{journal}{Geol. Mijnbouw (Leiden)} \textbf{\bibinfo{volume}{44}},
  \bibinfo{pages}{22} (\bibinfo{year}{1965}).

\bibitem[{\citenamefont{Duringer}(1988)}]{duringer}
\bibinfo{author}{\bibfnamefont{P.}~\bibnamefont{Duringer}},
  \bibinfo{type}{Th\`ese de doctorat dÕetat}, \bibinfo{school}{Universit\'e
  Louis Pasteur} (\bibinfo{year}{1988}).

\bibitem[{\citenamefont{Sneed and Folk}(1958)}]{sneed}
\bibinfo{author}{\bibfnamefont{E.~D.} \bibnamefont{Sneed}} \bibnamefont{and}
  \bibinfo{author}{\bibfnamefont{R.~L.} \bibnamefont{Folk}},
  \bibinfo{journal}{Journal of Geology} \textbf{\bibinfo{volume}{66}},
  \bibinfo{pages}{114} (\bibinfo{year}{1958}).

\bibitem[{\citenamefont{Illenberger}(1991)}]{illenberger}
\bibinfo{author}{\bibfnamefont{W.~K.} \bibnamefont{Illenberger}},
  \bibinfo{journal}{Journal of Sedimentary Petrology}
  \textbf{\bibinfo{volume}{61}}, \bibinfo{pages}{756} (\bibinfo{year}{1991}).

\bibitem[{\citenamefont{Benn and Ballantyne}(1992)}]{benn}
\bibinfo{author}{\bibfnamefont{D.~I.} \bibnamefont{Benn}} \bibnamefont{and}
  \bibinfo{author}{\bibfnamefont{C.~K.} \bibnamefont{Ballantyne}},
  \bibinfo{journal}{Journal of Sedimentary Petrology}
  \textbf{\bibinfo{volume}{62}}, \bibinfo{pages}{1147} (\bibinfo{year}{1992}).

\bibitem[{\citenamefont{Howard}(1992)}]{howard}
\bibinfo{author}{\bibfnamefont{J.~L.} \bibnamefont{Howard}},
  \bibinfo{journal}{Sedimentology} \textbf{\bibinfo{volume}{39}},
  \bibinfo{pages}{471} (\bibinfo{year}{1992}).

\bibitem[{\citenamefont{Hofmann}(1994)}]{hofmann}
\bibinfo{author}{\bibfnamefont{H.~J.} \bibnamefont{Hofmann}},
  \bibinfo{journal}{Journal of Sedimentology Research}
  \textbf{\bibinfo{volume}{64}}, \bibinfo{pages}{916} (\bibinfo{year}{1994}).

\bibitem[{\citenamefont{Graham and Midgley}(2000)}]{graham}
\bibinfo{author}{\bibfnamefont{D.~J.} \bibnamefont{Graham}} \bibnamefont{and}
  \bibinfo{author}{\bibfnamefont{N.~G.} \bibnamefont{Midgley}},
  \bibinfo{journal}{Earth Surface Processes and Landforms}
  \textbf{\bibinfo{volume}{25}}, \bibinfo{pages}{1473} (\bibinfo{year}{2000}).

\bibitem[{\citenamefont{Ehrlich and Weinberg}(1970)}]{ehrlich}
\bibinfo{author}{\bibfnamefont{R.}~\bibnamefont{Ehrlich}} \bibnamefont{and}
  \bibinfo{author}{\bibfnamefont{B.}~\bibnamefont{Weinberg}},
  \bibinfo{journal}{Journal of Sedimentary Petrology}
  \textbf{\bibinfo{volume}{40}}, \bibinfo{pages}{205} (\bibinfo{year}{1970}).

\bibitem[{\citenamefont{Clark}(1981)}]{clark}
\bibinfo{author}{\bibfnamefont{M.~W.} \bibnamefont{Clark}},
  \bibinfo{journal}{Journal of the International Association for Mathematical
  Geology} \textbf{\bibinfo{volume}{13}}, \bibinfo{pages}{303}
  (\bibinfo{year}{1981}).

\bibitem[{\citenamefont{Diepenbroek et~al.}(1992)\citenamefont{Diepenbroek,
  Bartholoma, and Ibbeken}}]{diepenbroek}
\bibinfo{author}{\bibfnamefont{M.}~\bibnamefont{Diepenbroek}},
  \bibinfo{author}{\bibfnamefont{A.}~\bibnamefont{Bartholoma}},
  \bibnamefont{and} \bibinfo{author}{\bibfnamefont{H.}~\bibnamefont{Ibbeken}},
  \bibinfo{journal}{Sedimentology} \textbf{\bibinfo{volume}{39}},
  \bibinfo{pages}{411} (\bibinfo{year}{1992}).

\bibitem[{\citenamefont{Bowman et~al.}(2001)\citenamefont{Bowman, Soga, and
  Drummond}}]{bowman}
\bibinfo{author}{\bibfnamefont{E.~T.} \bibnamefont{Bowman}},
  \bibinfo{author}{\bibfnamefont{K.}~\bibnamefont{Soga}}, \bibnamefont{and}
  \bibinfo{author}{\bibfnamefont{W.}~\bibnamefont{Drummond}},
  \bibinfo{journal}{Geotechnique} \textbf{\bibinfo{volume}{51}},
  \bibinfo{pages}{545} (\bibinfo{year}{2001}).

\bibitem[{\citenamefont{Drolon et~al.}(2000)\citenamefont{Drolon, Druaux, and
  Faure}}]{drolon}
\bibinfo{author}{\bibfnamefont{H.}~\bibnamefont{Drolon}},
  \bibinfo{author}{\bibfnamefont{F.}~\bibnamefont{Druaux}}, \bibnamefont{and}
  \bibinfo{author}{\bibfnamefont{A.}~\bibnamefont{Faure}},
  \bibinfo{journal}{Pattern Recognition Letters} \textbf{\bibinfo{volume}{21}},
  \bibinfo{pages}{473} (\bibinfo{year}{2000}).

\bibitem[{\citenamefont{Weisstein}(1999)}]{weisstein}
\bibinfo{author}{\bibfnamefont{E.~W.} \bibnamefont{Weisstein}},
  \emph{\bibinfo{title}{The CRC concise encyclopedia of mathematics}}
  (\bibinfo{publisher}{CRC Press}, \bibinfo{address}{New York},
  \bibinfo{year}{1999}).

\bibitem[{\citenamefont{Press et~al.}(1992)\citenamefont{Press, Flannery,
  Teukolsky, and Vetterling}}]{nr}
\bibinfo{author}{\bibfnamefont{W.~H.} \bibnamefont{Press}},
  \bibinfo{author}{\bibfnamefont{B.~P.} \bibnamefont{Flannery}},
  \bibinfo{author}{\bibfnamefont{S.~A.} \bibnamefont{Teukolsky}},
  \bibnamefont{and} \bibinfo{author}{\bibfnamefont{W.~T.}
  \bibnamefont{Vetterling}}, \emph{\bibinfo{title}{Numerical Recipes in C}}
  (\bibinfo{publisher}{Cambridge University Press}, \bibinfo{address}{New
  York}, \bibinfo{year}{1992}), \bibinfo{edition}{2nd} ed.

\bibitem[{nih()}]{nih}
\bibinfo{note}{NIH Image, public domain software available at
  http://rsb.info.nih.gov/nih-image/.}

\end{thebibliography}

\end{document}